\documentclass[%
reprint,
superscriptaddress,
amsmath,amssymb,
aps,
pra,
]{revtex4-2}

\usepackage{graphicx}
\usepackage{dcolumn}
\usepackage{bm}
\usepackage[colorlinks=true,citecolor=blue,linkcolor=red,urlcolor=blue]{hyperref}
\usepackage{comment}
\def\bra#1{\mathinner{\langle{#1}|}}
\def\ket#1{\mathinner{|{#1}\rangle}}
\def\braket#1{\mathinner{\langle{#1}\rangle}}

\def\proj#1{\ket{#1}\bra{#1}}

\def\tD{{\tilde{\Delta}}}

\begin{document}
\date{\today}

\title{Spin-photon coupling using circular double quantum dots}

\author{Ferdinand Omlor}
\affiliation{Division of Solid State Physics and NanoLund, Lund University, S-22100 Lund, Sweden}

\author{Florinda Viñas Boström}
\affiliation{Center for Quantum Devices, Niels Bohr Institute, University of Copenhagen, DK-2100 Copenhagen, Denmark}

\author{Martin Leijnse}
\affiliation{Division of Solid State Physics and NanoLund, Lund University, S-22100 Lund, Sweden}

\begin{abstract}
We propose and analyze a microwave spin-photon interface based on a circular double quantum dot, inspired by recent experimental observations of anisotropic g-factors and ring states in InAs nanowires. We develop an effective theoretical model capturing the interplay of spin-orbit coupling and the magnetic flux through the ring and show how ring states form at crossings of odd and even geometrical parity orbital states. Similar to bonding and antibonding states of conventional double quantum dots, the ring eigenstates can be changed into single dot states by detuning the dots, which enables a high degree of control over the system's properties.
Applying a tilted magnetic field induces spin-charge hybridization which enables spin-photon coupling. For low disorder, the photons couple states of simultaneously (almost) opposite spin and angular momentum. With increasing disorder, the spin-photon coupling becomes analogous to the flopping mode mechanism of conventional double quantum dots where the spin is hybridized with the bonding and antibonding orbital states without angular momentum.
We show that the system exhibits a second-order charge-noise sweet spot at a specific magnetic field angle, which lowers the system's sensitivity to dephasing while retaining a substantial spin-photon coupling strength.
Moreover, the photon coupling mechanism can be switched off either electrically, by detuning to the single-dot regime, or magnetically, by rotating the field to disable the spin-charge hybridization.
\end{abstract}

\maketitle

\section{Introduction \label{sec:intro}}

Semiconductor quantum dots (QDs) provide a versatile platform for engineering interactions between localized electronic states and the quantized electromagnetic field \cite{burkard_superconductorsemiconductor_2020}. Coupling QDs to microwave photons in superconducting resonators has enabled a range of cavity quantum electrodynamics experiments, reaching from strong electric dipole coupling with charge qubits \cite{mi_strong_2017, stockklauser_strong_2017} to photon-assisted tunneling \cite{kouwenhoven_observation_1994, liu_photon_2014} and dispersive readout schemes \cite{delbecq_coupling_2011, frey_dipole_2012}. Charge–photon coupling can readily reach the strong-coupling regime because the charge degree of freedom naturally carries a large electric dipole moment. However, charge qubits suffer from fast dephasing due to electrical noise, which severely limits their usefulness as quantum information carriers.

A natural path forward is to encode quantum information in the electron spin \cite{loss_quantum_1998, burkard_semiconductor_2023}. The spin degree of freedom in QDs is attractive due to long intrinsic coherence times and compatibility with established semiconductor fabrication technology. Unfortunately, the direct magnetic-dipole interaction between a single spin and cavity photons is inherently weak \cite{RASHBA1991131}, making coherent spin–photon interfaces challenging to realize. To overcome this, hybridization of spin and charge degrees of freedom using spin-orbit coupling (SOC) or a magnetic field gradient can be used \cite{rashba1960properties, pekar_1965, nowack_coherent_2007, trif_spin_2008, hu_strong_2012, benito_input-output_2017, liu_spin-orbit_2018}. This hybridization enhances the effective dipole moment of the spin qubit and facilitates coupling to a cavity’s electric field. 
Using this, coherent spin-photon interfaces \cite{samkharadze_strong_2018, mi_coherent_2018, landig_coherent_2018, ungerer_strong_2024, burkard_superconductorsemiconductor_2020}, dispersive spin readout \cite{kam_fast_2024} as well as long-range spin-spin gates \cite{borjans_resonant_2020, harvey-collard_coherent_2022,  dijkema_cavity-mediated_2025} have been demonstrated during the last years.

While spin-charge hybridization is crucial for spin-photon coupling, it comes at a cost: 
the effective electronic dipole moment of the spin qubit also makes it susceptible to fluctuations of the electrostatic environment, i.e., charge noise, which decreases the coherence time \cite{chirolli_decoherence_2008, kuhlmann_charge_2013, burkard_semiconductor_2023}. As a result, a key objective in the design of spin-photon devices is to find ``sweet spots"—parameter regimes where the system is robust against fluctuations of the electric field. Recent theoretical and experimental efforts have found such regimes by fine-tuning the system parameters \cite{benito_electric-field_2019, croot_flopping-mode_2020, ungerer_dephasing_2025}.

\begin{figure}
    \centering
    \includegraphics[width=6cm]{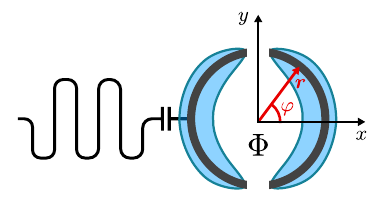}
    \caption{
    A circular DQD (blue) spin-photon interface.
    We model the system using an effective one-dimensional ring (dark gray) with two barriers such that the electron position is given by $\varphi$. The electrons can tunnel through the barriers and the spin-orbit interaction causes the formation of ring states for specific inter-dot detunings. For the ring states, the magnetic flux $\Phi$ through the ring acts as an orbital contribution to the electron $g$-factor. The cavity is capacitatively coupled to one of the QDs and can be used to drive transitions between the electronic spin states using a flopping-mode mechanism.} 
    \label{fig:ring_dqd}
\end{figure}

In this work, we study a circular double quantum dot (DQD) system, as displayed in Fig.~\ref{fig:ring_dqd}. This idea is inspired by recent experimental observations of ring states with highly anisotropic $g$-factors in thin crystal-phase-defined DQDs in InAs nanowires \cite{potts_electrical_2019, potts_symmetry-controlled_2021, debbarma_effects_2022, debbarma_josephson_2022}, but can be realized in any system which has significant SOC and where circular DQDs can be formed. 
Reference~\cite{potts_electrical_2019} showed how the SOC creates ring states at crossings of even and odd orbital parity states. These states have a finite angular momentum and therefore high orbital contributions to the $g$-factor, leading to relatively large splittings for low magnetic field strengths.

In this work, we derive an effective analytic model for this system and show that by applying a tilted magnetic field, spin-charge hybridization can be induced which allows spin-photon coupling. Moreover, for low disorder, photons couple electron states of simultaneously (almost) opposite spin and angular momentum. 
Crucially, we identify a second-order sweet spot in the system’s parameter space, where both first and second derivatives of the level splitting vanish for fluctuations of the electrostatic environment. Therefore, dephasing due to charge noise is suppressed while a substantial spin-photon coupling strength is retained. Furthermore, the coupling can be switched off electrically or magnetically, providing a high degree of tunability. These results point to a promising pathway for tunable and low-decoherence spin-photon interfaces in semiconductor QD systems.

This paper is organized as follows:
The model section~\ref{sec:model} starts with the derivation of a one-dimensional ring Hamiltonian, as displayed in Fig.~\ref{fig:ring_dqd}, in Sec.~\ref{sec:ring_H} and introduces the potential which defines the DQDs in Sec.~\ref{sec:step_pot}. We derive the cavity coupling mechanism in Sec.~\ref{sec:cavity_coupl}.
The result section~\ref{sec:results} starts with the analytic solution of the orbital zero-field Hamiltonian in Sec.~\ref{sec:orb0field}.  In Sec.~\ref{sec:eff_cross_H} we show how ring states emerge for crossings of even and odd geometrical parity states and derive an effective Hamiltonian describing these. The resulting energy spectrum is discussed in Sec.~\ref{sec:spectrum} while Sec.~\ref{sec:2lvl} derives analytical expressions for the energy splittings and Rabi coupling strengths in the case of small magnetic field strength. In Sec.~\ref{sec:sweet_spot}, we show the existence of a second-order charge noise sweet spot and
in Sec.~\ref{sec:InAs}, we use the experimental data in Ref.~\cite{potts_electrical_2019} to estimate the achievable spin-photon coupling strength for that particular system and discuss how this can be improved. Section~\ref{sec:conclusion} summarizes and concludes this work.

\section{Model} \label{sec:model}

This section lays the foundation on which the results are built by defining the ring Hamiltonian, the DQD potential and the microwave resonator coupling mechanism.

\subsection{The Ring Hamiltonian} \label{sec:ring_H}

The starting point for our model is non-interacting electrons described by the Hamiltonian
\begin{align}
    H &= H_{\mathrm{orb}} + H_\mathrm{SOC} + H_Z, \label{eq:Ham}
\end{align}
where $H_{\mathrm{orb}}$ is the orbital Hamiltonian, given by
\begin{align}
    H_{\mathrm{orb}} = \frac{1}{2m^*} \left( -i\hbar \bm{\nabla} + e\bm{A}(\bm{r}) \right)^2 + V_\mathrm{3D}(\bm{r}),
\end{align}
with the electron effective mass $m^*$, charge $e$ and vector potential $\bm{A}(\bm{r})$. We assume that the potential $V_\mathrm{3D}(\bm{r}) = V(\varphi) V_z(z)V_r(r)$ confines the electrons to a one-dimensional ring of radius $R$ as displayed in Fig.~\ref{fig:ring_dqd}. The vector potential for an arbitrary magnetic field $\bm{B}$ in the Coulomb gauge is then given by
\begin{align}
    \bm{A} (R, \varphi) = \frac{R B_z}{2} \bm{e}_\varphi + R [\sin(\varphi)B_x - \cos(\varphi)B_y] \bm{e_z},
\end{align}
and the one-dimensional orbital Hamiltonian is
\begin{align} 
    H_\mathrm{orb}^\mathrm{1D} &= \epsilon_R \left(\frac{L}{\hbar} - \frac{\Phi}{\Phi_0}\right)^2+ V(\varphi) \notag \\
    &+ 4\epsilon_R\frac{A_\circ^2}{\Phi_0^2}\left[\sin(\varphi)B_x  - \cos(\varphi) B_y\right]^2, \label{eq:Horb}
\end{align}
where the angular momentum operator is given by $L = -i\hbar \partial_\varphi$. The magnetic flux through the ring is given by $\Phi = \pi R^2 B_z$ and  $\Phi_0 = -\frac{2\pi\hbar}{e}$ is the magnetic flux quantum. $A_\circ = \pi R^2$ denotes the area enclosed by the ring and $\epsilon_R = \frac{\hbar^2}{2m^*R^2}$ is the orbital energy scale. In addition to the Aharonov-Bohm term $\propto \Phi/\Phi_0$, the vector potential gives rise to a $\varphi$-dependent term which couples the orbital wavefunction to the in-plane magnetic field ($B_x, B_y$) \cite{planelles_quantum_2006}.

The second term of $H$ describes a Rashba type SOC given by $H_\mathrm{SOC} = \frac{\alpha_R}{\hbar} (\bm{e} \times \bm{p})\cdot \bm{\sigma}$, where $\alpha_R$ is the Rashba SOC strength and $\bm{e}$ is a unit vector in the direction of the electric field from which the SOC originates \cite{rashba1960properties, manchon_new_2015}. $\bm{p}$ is the canonical momentum of the electron and $\bm{\sigma}$ a vector of Pauli spin operators. We consider the case of the electric field pointing in the radial direction which leads to the SOC Hamiltonian
\begin{align}
    H_\mathrm{SOC} &= \frac{\alpha_R}{\hbar} p_\varphi \sigma_z \\
    &= \alpha \frac{L}{\hbar} \sigma_z + \frac{1}{2} \mu_B g_z^\mathrm{SOC} \cdot  B_z \sigma_z, \label{eq:SOC}
\end{align}
where $\mu_B = \frac{e\hbar}{2m}$ is the Bohr magneton and $\alpha = \frac{\alpha_R}{R}$. The second term is an anisotropic correction to the $g$-factor given by $g_z^\mathrm{SOC} = \frac{\alpha}{\epsilon_R}$. In the following, we assume that the SOC is small compared to the orbital energy scale, i.e., $\epsilon_R \gg \alpha_R/R$ and therefore we neglect $g_z^\mathrm{SOC}$.
Reference \cite{meijer_one-dimensional_2002} derived the ring SOC Hamiltonian for the alternative case of $\bm{e}$ pointing in the $z$-direction.

Finally,
\begin{align}
    H_Z = \frac{1}{2}\mu_B g \bm{B}\cdot \bm{\sigma}, \label{eq:Zeeman}
\end{align} 
denotes the Zeeman Hamiltonian with the effective spin $g$-factor $g$.

\subsection{DQD Ring Potential} \label{sec:step_pot}

\begin{figure}
    \centering
    \includegraphics[width=8.7cm]{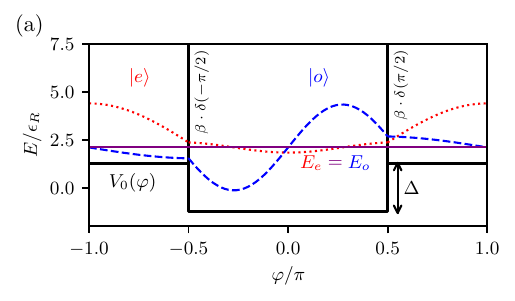}
    \includegraphics[width=8.7cm]{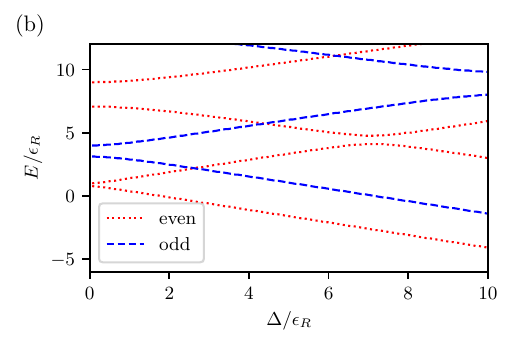}
    \caption{(a) Circular potential $V_0(\varphi)$ (black) of the ring DQD. The two QDs are separated by two Dirac barriers $\beta \cdot \delta(\pm\pi/2)$ and detuned by $\Delta$. For the chosen parameters, $\beta/\epsilon_R = 7$ and $\Delta/\epsilon_R = 2.62$, the even $\ket{e}$ and odd $\ket{o}$ eigenstate are degenerate $E_e = E_o$. The wavefunctions $\braket{\varphi|e}$ (red dotted line) and $\braket{\varphi|o}$ (blue dashed line) are plotted in arbitrary units. (b) Energy spectrum of the odd and even eigenstates as a function of $\Delta$. }
    \label{fig:H0_spectrum}
\end{figure}

We describe the circular DQD using an azimuthal potential $V_0(\varphi)$ as a function of the angle $\varphi \in (-\pi, \pi]$ shown in Fig.~\ref{fig:ring_dqd} and defined by
\begin{align}
V_0(\varphi) &=  \delta(\varphi \pm \pi/2) \cdot \beta \notag  \\ 
&+\begin{cases}
    -\frac{\Delta}{2}, & -\pi/2 < \varphi \le \pi/2 \\
    \frac{\Delta}{2}, & \text{otherwise}.
    \end{cases}  \label{eq:V0}
\end{align}
The DQD is modeled by $V_0(\varphi)$ being constant within each QD [see Fig.~\ref{fig:H0_spectrum}(a)],
while barriers between the QDs are introduced by two Dirac distributions of strength $\beta$ and $\Delta$ defines the detuning of the QDs.
We also introduce a small disorder potential $V_\delta \ll \epsilon_R$, such that the complete potential is given by $V(\varphi) = V_0(\varphi) + V_\delta(\varphi)$.
$V_0(\varphi)$ is parity-symmetric under the operation $\Pi:\varphi \rightarrow -\varphi$. This symmetry is broken by the disorder term. We assume that this symmetry breaking is ``small" and can be treated perturbatively.

We stress that the following approach also works for more sophisticated potentials $V_0$, e.g., with finite width barriers or non-uniform detuning of the QDs, as long as the parity symmetry $\Pi$ is satisfied for $V_0$ and $V_\delta$ can be treated perturbatively. An analytical solution for finite width barriers is given in Ref.~\cite{costa_electronic_2017}.

\subsection{Cavity Coupling} \label{sec:cavity_coupl}

We consider a single-mode microwave cavity which is capacitively coupled to one of the QDs as displayed in Fig.~\ref{fig:ring_dqd}. Realistically, a microwave cavity couples to both QDs, but we assume that the coupling is stronger to the closer QD and ignore an irrelevant simultaneous shift of both QD potentials. In this treatment, the microwave cavity effectively couples to the inter-QD detuning and we substitute $\Delta$ in $V_0(\varphi)$, Eq.~\eqref{eq:V0}, with
\begin{align} 
    \Delta = \Delta_0 + \epsilon_\gamma (a^\dagger + a), \quad \epsilon_\gamma = \alpha_g \hbar \omega \sqrt{\frac{\pi Z}{R_Q}}\,, \label{eq:res_defs}
\end{align}
where $a$ is the photon annihilation operator of a single mode resonator with frequency $\omega$ \cite{blais_circuit_2021, childress_mesoscopic_2004}. Furthermore, $\Delta_0$ is a constant inter-dot bias, $Z$ the impedance of the resonator, $R_Q = h/e^2$ the resistance quantum and $\alpha_g$ the coupling lever-arm. Inserting this into the system Hamiltonian $H$, Eq.~\eqref{eq:Ham}, and expanding to first order in $(a^\dagger + a)$, results in the Rabi Hamiltonian
\begin{align} 
    H_\omega &= \hbar\omega a^\dagger a + \sum_n E_n \proj{n} \notag \\
    &+ \sum_{m\ne n} (\gamma_x^{mn} \mathcal{X}_{mn} + \gamma_y^{mn} \mathcal{Y}_{mn} + \gamma_z^{mn} \mathcal{Z}_{mn})(a^\dagger + a), \label{eq:H_omega}
\end{align}
where $E_n$ are the eigenvalues corresponding to the respective eigenstates $\ket{n}$ of the unperturbed Hamiltonian $H$ for $\Delta = \Delta_0$. Here we neglect first-order terms of $(a^\dagger + a)$ causing an overall energy shift, and add the cavity energy term $\hbar \omega a^\dagger a$.
$\mathcal{X}_{mn}, \mathcal{Y}_{mn}$ and $\mathcal{Z}_{mn}$ denote Pauli operators between the eigenstates $\ket{m}$ and $\ket{n}$ and the coupling rates are given by
\begin{align}
    \gamma_x^{mn} &= \frac{\epsilon_\gamma}{2} \left(\braket{m|H'|n} + \braket{n|H'|m}\right), \label{eq:gamma_x} \\
    \gamma_y^{mn} &= \frac{\epsilon_\gamma}{2i} \left(\braket{m|H'|n} - \braket{n|H'|m}\right), \label{eq:gamma_y} \\
    \gamma_z^{mn} &= \frac{\epsilon_\gamma}{2} \left(\braket{m|H'|m} -  \braket{n|H'|n}\right), \label{eq:gamma_z}
\end{align}
where $\braket{m|H'|n} = \braket{m|(\partial_\Delta H)\big|_{\Delta = \Delta_0}|n}$. The transversal couplings $\gamma_x^{mn}$ and $\gamma_y^{mn}$ can be used to drive transitions between the states, as well as for dispersive readout of them \cite{blais_circuit_2021}.
On the other hand, $\gamma_z^{mn}$ describes longitudinal coupling of the 
cavity mode to the DQD states. Due to the DQD state-dependent ($\mathcal{Z}_{mn}$) shift of the cavity canonical coordinate $\propto a + a^\dagger$, this coupling can also be used to read out the states \cite{kerman_quantum_2013, didier_fast_2015, harpt_ultra-dispersive_2025}, as well as for two-qubit entanglement operations \cite{kerman_quantum_2013}.
However, in the limit of $|\gamma_{z}^{mn}| \ll |E_m-E_n|$, the longitudinal coupling has only a small effect on the level spacing which averages out if in addition $|\gamma_{z}^{mn}| \ll\omega$. Moreover, if $|\gamma_{x,y}^{mn}| \ll |E_m-E_n|$, a rotating wave approximation can be deployed to arrive at the well-known Jaynes-Cummings Hamiltonian \cite{jaynes_comparison_1963} for the transition $\ket{m} \leftrightarrow \ket{n}$ 
\begin{align}
    H_{JC}^{mn} = \hbar \omega  a^\dagger a + \frac{\Omega^{mn}}{2} \mathcal{Z}_{mn} + \gamma_\perp^{mn} \left(a^\dagger\ket{n}\!\bra{m} + a\ket{m}\!\bra{n}\right),
\end{align}
where $\Omega^{mn} = E_m -E_n$ and $\gamma_\perp^{mn} = \sqrt{(\gamma_x^{mn})^2 + (\gamma_y^{mn})^2}$.

\section{Results} \label{sec:results}

\subsection{Orbital Zero Field Solution} \label{sec:orb0field}

In this section, we find an analytical solution for the unperturbed, zero-field Hamiltonian
\begin{align}
    H_0 = \epsilon_R \frac{L^2}{\hbar^2} + V_0(\varphi), \label{eq:H0}
\end{align}
which is equal to $H_\mathrm{orb}^\mathrm{1D}$ in Eq.~\eqref{eq:Horb} in the case of $\bm{B} = 0$ and for $V_\delta(\varphi) = 0$. 
Using the parity symmetry of $V_0(\varphi)$, the eigenstates $\ket{n}$ of $H_0$ must be even  or odd under the symmetry transformation $\Pi$. 
To solve the Schrödinger equation, we use the ansatz
\begin{align}
    n \mathrm{~even}:~&\braket{\varphi|n} = \begin{cases}
    A_n \cos(k^+_n(\varphi + \pi)) & -\pi < \varphi \le -\frac{\pi}{2} \\
    B_n \cos(k^-_n\varphi) & -\frac{\pi}{2}< \varphi \le \frac{\pi}{2} \\
    A_n \cos(k^+_n(\varphi - \pi)) & \frac{\pi}{2} < \varphi \le \pi
    \end{cases}, \label{eq:wavefunc_even}\\
    n \mathrm{~odd}:~&\braket{\varphi|n}= \begin{cases}
    C_n \sin(k^+_n(\varphi + \pi)) & -\pi < \varphi \le -\frac{\pi}{2} \\
    D_n \sin(k^-_n\varphi) & -\frac{\pi}{2} < \varphi \le \frac{\pi}{2} \\
    C_n \sin(k^+_n(\varphi - \pi)) & \frac{\pi}{2} < \varphi \le \pi
    \end{cases}, \label{eq:wavefunc_odd}
\end{align}
with wave numbers $k^\pm_n = \sqrt{(E_n\pm\Delta/2)/\epsilon_R}$ and eigenenergies $E_n$. 
By applying continuity conditions for the wavefunctions and their derivatives at $\varphi = \pm \pi/2$, the coefficients $A_n, B_n, C_n, D_n$ and eigenenergies $E_n$ are found, see Appendix \ref{sec:cont_cond}.
The resulting energy spectrum is plotted in Fig.~\ref{fig:H0_spectrum}(b).
As one can see from the ansatz in Eqs.~\eqref{eq:wavefunc_even}, \eqref{eq:wavefunc_odd}, the states are standing waves with $\braket{L} = 0$.
At avoided crossings of same-parity states, the electron forms bonding and anti-bonding molecular states; while away from these, the electron is mainly localized on one of the QDs. For barrier strengths $\beta > \epsilon_R$, this is also the case at the crossings of different parity states, see Fig.~\ref{fig:H0_spectrum}(a). Therefore, we can view the even and odd state at a crossing as localized single QD states.

\subsection{Effective Crossing Hamiltonian} \label{sec:eff_cross_H}

In the following, we focus on crossings of odd and even states and derive an effective Hamiltonian which accounts for the SOC, the small disorder $V_\delta$ and the effect of a magnetic field. 
To simplify this problem, we assume that an in-plane magnetic field is aligned with the $x$-axis and therefore the term $\propto \sin(x)^2 B_x^2$ of $H_\mathrm{orb}$ [See Eq.~\eqref{eq:Horb}] does not break the parity symmetry $\Pi$ and mainly causes a shift of the crossing states without coupling them. Additionally, we assume that the magnetic flux through the ring is small compared to the flux quantum, $\Phi \ll \Phi_0$, and therefore the 1D orbital Hamiltonian $H_\mathrm{orb}^\mathrm{1D}$, Eq.~\eqref{eq:Horb}, is approximated by
\begin{align}
    H_\mathrm{orb}^\mathrm{1D} \approx H_0 - \mu_B^*B_z \frac{L}{\hbar} + 4\epsilon_R\frac{A_\circ^2}{\Phi_0^2}B_x^2 \sin(\varphi)^2 ,
\end{align}
where $\mu_B^* = \frac{e\hbar}{2m^*}$ is the Bohr magneton using the effective electron mass. 

Now, the complete Hamiltonian [which also includes SOC , Eq.~\eqref{eq:SOC}, and the Zeeman term, Eq.~\eqref{eq:Zeeman}] is given by
\begin{align}
    H &= H_0 + V_\delta(\varphi) + (\alpha \sigma_z- \mu_B^*B_z )\frac{L}{\hbar} \notag\\ 
    &+ \frac{1}{2}\mu_B g \bm{B}\cdot \bm{\sigma} + 4\epsilon_R\frac{A_\circ^2}{\Phi_0^2}B_x^2 \sin(\varphi)^2. 
\end{align}

Except for $H_0$, we assume that all terms of $H$ are small compared to the orbital energy $\epsilon_R$. Therefore, for $\Delta$ close to a crossing of an even and an odd parity orbital state, $\ket{e}$ and $\ket{o}$, we can restrict the Hilbert space to the crossing states and introduce all additional terms to $H_0$ in first-order perturbation theory.

Using this, we investigate the angular momentum matrix elements and find $\braket{e|L|e} = -i\hbar \braket{e|\partial_\varphi|e} = 0 = \braket{o|L|o}$, because of the parity symmetry of $\ket{e}$ and $\ket{o}$. However, the matrix elements between different parity states are in general $\braket{e|L|o}=- i l \hbar \neq 0$ with $l \in \mathbb{R}$. Notably, $l$ is in general not an integer and highly depends on the spread of $\ket{e}$ and $\ket{o}$ over both dots and therefore the barrier strength.

Next, the disorder potential $V_\delta$ is also treated using first-order perturbation theory. We divide the disorder potential into an even and an odd part regarding the parity symmetry $\Pi$
\begin{align}
    V_\delta(\varphi) = V_\delta^e(\varphi) +  V_\delta^o(\varphi). \label{eq:Vdelta}
\end{align}
The even part $V^e_\delta$ introduces an energy shift to the odd and even eigenstates $\braket{e|V_\delta^e|e} \ne 0$ and $\braket{o|V_\delta^e|o} \ne 0$ while $\braket{e|V_\delta^e|o} = 0$. The odd part $V^o_\delta$ breaks the parity symmetry and introduces an off-diagonal disorder coupling $\braket{e|V_\delta^o|o} = \braket{o|V_\delta^o|e} = \delta/2$ while $\braket{e|V_\delta^o|e}  = \braket{o|V_\delta^o|o} = 0$.

Finally, the in-plane orbital magnetic term $\propto B_x^2$ introduces matrix elements $\propto \braket{e|\sin(\varphi)^2|e},\braket{o|\sin(\varphi)^2|o}$ while $\braket{e|\sin(\varphi)^2|o} = 0$.
For simplicity, we assume that $\braket{e|\sin(\varphi)^2|e} \approx\braket{o|\sin(\varphi)^2|o}$ and neglect the overall shift of all states. With this, we arrive at the effective Hamiltonian
\begin{align} 
    H_{\mathrm{eff}} &= \frac{\tD}{2} \Sigma_z + \frac{\delta}{2} \Sigma_x - l\mu_B^* B_z \Sigma_y \notag \\  &+ l\alpha \sigma_z \Sigma_y
    + \mu_B \frac{g}{2} \bm{B} \cdot \bm{\sigma}, \label{eq:H_4lvl}
\end{align}
where $\tD = E_e - E_o$ is the energy difference of the unperturbed states $\ket{e}, \ket{o}$, as derived in Sec.~\ref{sec:orb0field}, and the orbital Pauli operators are given by
\begin{align}
    &\Sigma_x = \ket{o}\!\bra{e} + \ket{e}\!\bra{o}, \\
    &\Sigma_y = i\ket{o}\!\bra{e} - i\ket{e}\!\bra{o}, \\
    &\Sigma_z = \ket{e}\!\bra{e} - \ket{o}\!\bra{o}.
\end{align}

The effective Hamiltonian in Eq.~\eqref{eq:H_4lvl} can be understood as a DQD Hamiltonian. As pointed out in Sec.~\ref{sec:orb0field}, $\ket{e}$ and $\ket{o}$ are mainly localized on a single QD, see Fig.~\ref{fig:H0_spectrum}(a). Therefore, the terms $\frac{\delta}{2} \Sigma_x$ and $-l\mu_B^* B_z \Sigma_y$ both describe spin-independent tunneling between the QDs. On the other hand, the SOC term $l\alpha \sigma_z \Sigma_y$ couples orbital and spin degrees of freedom and can be understood as a spin-dependent tunneling process.

Furthermore, the Pauli operators $\Sigma_{x,y,z}$ give an intuitive interpretation of the electronic orbital:
The localized single QD states are eigenstates of $\Sigma_z$. Eigenstates of $\Sigma_x$ resemble bonding and antibonding DQD orbitals with overall $\braket{L} = 0$. But different from conventional DQDs, an eigenstate of $\Sigma_x$ is bonding and antibonding at once; at one of the barriers, the wavefunctions interfere constructively while at the other destructively, see Fig.~\ref{fig:H0_spectrum}(a). Therefore, breaking of the parity symmetry $\Pi$ introduces a splitting term of these states, $ \frac{\delta}{2} \Sigma_x$, to $H_\mathrm{eff}$.
Ring states, on the other hand, are eigenstates of $\Sigma_y$. This can be seen by formulating the angular momentum operator in the crossing state subspace, $L = \hbar l \, \Sigma_y$. Therefore, in eigenstates of $\Sigma_y$, the electron travels around the ring and picks up the Aharonov-Bohm phase caused by the magnetic flux through the ring. 

To describe the coupling to a microwave resonator,
we assume that the resonator effectively couples to the detuning $\Delta$ and derive the Rabi Hamiltonian, as described in Sec.~\ref{sec:cavity_coupl}.
The coupling strengths of the Rabi Hamiltonian, Eq.~\eqref{eq:H_omega}, are obtained by calculating the matrix elements of
$\partial_\Delta H_\mathrm{eff} = \zeta \Sigma_z/2$ where $\zeta = \partial_\Delta \tD$. The transversal Rabi coupling strength between two eigenstates $\ket{m},\ket{n}$ is given by
\begin{align}
    \gamma_\perp^{mn} = \sqrt{(\gamma_x^{mn})^2 + (\gamma_y^{mn})^2} = \frac{1}{2}\epsilon_\gamma \zeta |\braket{m|\Sigma_z|n}|, \label{eq:long_coupl}
\end{align}
where $\epsilon_\gamma$ is defined in Eq.~\eqref{eq:res_defs} and, because $\Sigma_z$ describes on which QD the electron is localized, $\braket{m|\Sigma_z|n}$ is analogous to the transition dipole moment between the states $\ket{m}$ and $\ket{n}$ used in atomic physics.
Furthermore, the longitudinal coupling strength between the states $\ket{m}$ and $\ket{n}$ is given by
\begin{align}
    \gamma_z^{mn} = \epsilon_\gamma \zeta \frac{\braket{m|\Sigma_z|m} - \braket{n|\Sigma_z|n}}{4} = \epsilon_\gamma \zeta \frac{\partial_\tD \Omega^{mn}}{2},\label{eq:trans_coupl}
\end{align}
where the energy splitting $\Omega^{mn} = E_m - E_n$.
\begin{figure*}[]
    \centering
    \includegraphics[height=4.6cm]{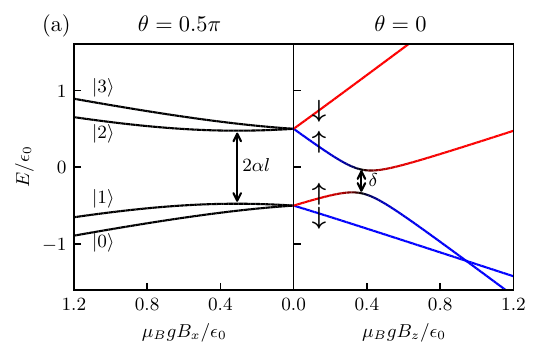}%
    \includegraphics[height=4.6cm]{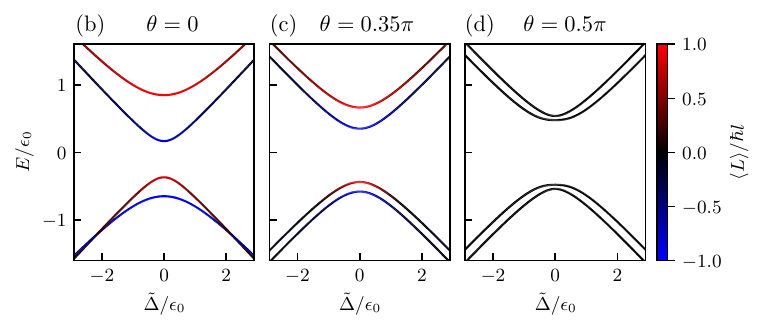}
    \caption{Crossing of an even and an odd orbital eigenstate. (a) Eigenenergies of $H_\mathrm{eff}$ as a function of the bare electron Zeeman energy $\mu_B g B_{x,z}$ for $\tD = 0$. In the left panel, $\bm{B}||\bm{e}_x$, i.e. $\theta = \arctan(B_x/B_z) = \pi/2$ and in the right panel $\bm{B}||\bm{e}_z$, $\theta = 0$.
    At $\bm{B} = 0$, the spectrum consists of two degenerate Kramers doublets which are split by $\epsilon_0 = \sqrt{4\alpha^2 l^2 + \delta^2}$. For $\theta = 0$, spin and orbit are separable and the spin is a good quantum number. Same-spin states undergo an avoided crossing at $B_z = \pm \alpha/\mu_B^*$ with splitting $\delta$. For $\theta = \pi/2$, spin and orbit are hybridized and $\braket{L} = 0$.
    We denote the states in numerical ascending energetic order. The states $\ket{1}$ and $\ket{2}$ have an avoided crossing with a gap of two times the SOC energy $2\alpha l$ at $\mu_B g B_x = \pm \delta/2$.
    (b), (c), (d) Eigenenergies of $H_\mathrm{eff}$ as a function of the detuning $\tD$ for $\mu_B g |\bm{B}| = 0.2 \epsilon_0$, for $\theta = 0$ (b), $\theta = 0.35\pi$ (c), $\theta = 0.5\pi$ (d). For $\theta = 0$ the splitting between the two lowest states is maximal for $\tD = 0$, whereas for $\theta = 0.5\pi$ it is minimal. Therefore a second-order sweet spot of the splitting can be found at $\theta = 0.35\pi$ where $\partial_{\tD} (E_1 - E_0) = \partial_{\tD}^2 (E_1 - E_0) = 0$ at $\tD = 0$. The system parameters used for the plots are given in Tab.~\ref{tab:system paramets}.}
    \label{fig:Heff_spectrum}
\end{figure*}
\begin{figure}[]
    \centering
    \includegraphics[width=8.7cm]{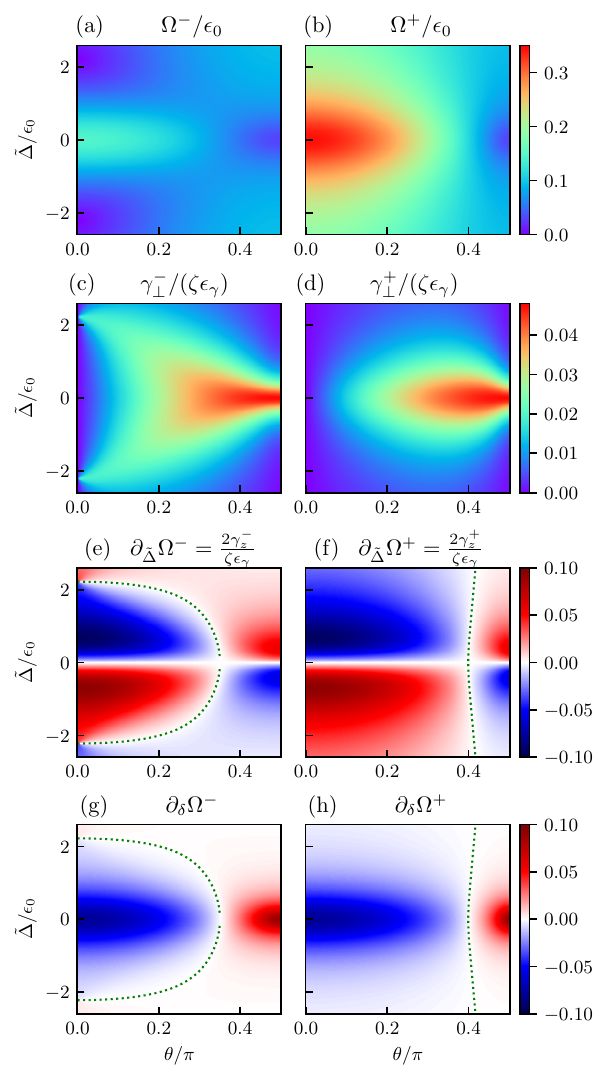}
    \caption{(a) Lower doublet splitting $\Omega^-$ and (b) upper doublet splitting $\Omega^+$ divided by the splitting of the upper and lower doublet at $B = 0$, $\epsilon_0 = \sqrt{4\alpha^2 l^2 + \delta^2}$, as a function of $\tD$ and the magnetic field angle $\theta = \arctan(B_x/B_z)$ for the constant field strength $\mu_B |g\bm{B}|/\epsilon_0 = 0.1$.
    (c) Rabi transversal coupling strength of the lower doublet $\gamma_\perp^-$ and (d) upper doublet $\gamma_\perp^+$ divided by $\zeta$ and the resonator coupling energy $\epsilon_\gamma$, Eq.~\eqref{eq:res_defs}.
    (e) First $\tD$-derivative of the  lower doublet splitting, $\partial_\tD\Omega^-$ and (f) upper doublet $\partial_\tD\Omega^+$. These also show the longitudinal Rabi coupling strength $\gamma_z^\pm$ using the relation $\partial_\tD\Omega^\pm = 2\gamma_z^\pm/(\zeta \epsilon_\gamma)$.
    (g) First $\delta$-derivative of the  lower doublet splitting, $\partial_\delta\Omega^-$ and (h) upper doublet $\partial_\delta\Omega^+$.
    The green dotted line in (e)-(h) shows the sweet spot angle $\theta_{ss}^\pm$ where $\partial_\tD\Omega^\pm = \partial_\delta\Omega^\pm = 0$.
    The system parameters used for the plots are given in Tab.~\ref{tab:system paramets}.}
    \label{fig:small_field}
\end{figure}
\begin{table}[]
    \centering
    \begin{tabular}{c|c|c|c|c|c}
         $g$& $m^*$ & $R$ & $\alpha_R$ & $l$ & $\delta/\epsilon_0$ \\ \hline
         $-12$& $0.026m$ & $17.5\,\mathrm{nm}$ & $2.024\,\mathrm{meV\cdot nm}$ & $0.4$ & $0.3$
    \end{tabular}
    \caption{System parameters used for the plots in Figs.~\ref{fig:Heff_spectrum},  \ref{fig:small_field} and \ref{fig:all_couplings}. The values of $g$, $m^*$ and $\alpha_R$ are inspired by InAs properties \cite{potts_electrical_2019}. The angular momentum number $l = -\braket{e|L|o}/(i\hbar)$ is obtained from the crossing of the second even and first odd states, shown in Fig.~\ref{fig:H0_spectrum}(a).
    The zero field splitting is $\epsilon_0 = \sqrt{4\alpha^2 l^2 + \delta^2}$.} 
    \label{tab:system paramets}
\end{table}

\subsection{Spectrum} \label{sec:spectrum}

For $\bm{B} = 0$, the four-fold orbital and spin degeneracy at $\tD=0$ of $H_0$, Eq.~\eqref{eq:H0}, is split into two Kramers doublets by the SOC and disorder coupling in $H_\mathrm{eff}$, Eq.~\eqref{eq:H_4lvl}. This results in an avoided crossing with a splitting between the Kramers doublets given by
\begin{align}
    \epsilon = \sqrt{\tD^2 + 4l^2 \alpha^2 + \delta^2}. \label{eq:epsilon}
\end{align}
In the following, we study the anisotropic effects of a magnetic field on the electronic spectrum.
Figure~\ref{fig:Heff_spectrum}(a) shows the splitting of the Kramers pair as a function of the magnetic field strength. We label the eigenstates $\ket{n}$ in energetically ascending order $n = 0,\dots, 3$.
For $\bm{B}||\bm{e}_z$ and $\tD = 0$, spin and orbital degrees of freedom are separable and the spin in the $z$-direction is a good quantum number.
The same-spin states $\ket{1}$ and $\ket{2}$ undergo an avoided crossing at $B_z = \alpha/\mu_B^*$ with a splitting $\delta$ which rotates the orbital quantization: whereas far away from the avoided crossing, i.e., $l(\mu_B^*B_z - \alpha)\gg \delta/2$, the states are eigenstates of $\Sigma_y$ with angular momentum $\braket{L} = \pm l \hbar$, at the crossing, $\ket{1}$ and $\ket{2}$ are eigenstates of $\Sigma_x$ with $\braket{L} = 0$.
For $\delta = 0$, the avoided crossing becomes degenerate and the angular momentum is a good quantum number for all $B_z$.
On the other hand, for $\bm{B}||\bm{e}_x$ and $\tD = 0$, spin and orbit are hybridized and all states have vanishing angular momentum $\braket{L} = 0$.
The doublet splitting tends towards $\delta$ with increasing $B_x$.

Interestingly, the magnetic-field dependence of the spectrum for $\tD=0$ is analogous to spin-valley systems such as carbon nanotubes \cite{laird_quantum_2015} and bilayer graphene \cite{banszerus_spin-valley_2021, kurzmann_kondo_2021, denisov_spinvalley_2025} where here the angular momentum $L$ takes the role of the valley.
However, in contrast to valley systems, where the valley is a degree of freedom on the crystal symmetry level, the angular momentum can be influenced macroscopically more easily. Like the bonding and anti-bonding DQD states $\propto \Sigma_x$, the angular momentum states $\propto \Sigma_y$ go over into single QD states $\propto \Sigma_z$ with vanishing $\braket{L}$ by increasing the detuning $\tD$.
This is displayed for increasing magnetic field angles $\theta = \arctan(B_x/B_z)$ and constant magnitude $\mu_B|g\bm{B}| = 0.2 \sqrt{4l^2 \alpha^2 + \delta^2}$ in Figures~\ref{fig:Heff_spectrum}(b)-(d). 

With increasing $\theta$, the angular momentum $\braket{L}$ of the states is reduced and for an in-plane magnetic field, $\theta = \pi/2$, $\braket{L} = 0$. In this configuration, the splitting is minimal at $\tD = 0$, which is caused by a (partial) suppression of the Zeeman splitting due to the SOC-induced spin-charge hybridization.
The change from a maximum at $\theta = 0$ to a minimum at $\theta = \pi/2$ of the upper and lower doublet splitting means that for directions in between, both doublets have a second-order sweet spot regarding fluctuations in $\tD$. There, the first and second derivatives of the lower (upper) doublet splitting vanish, $\partial_\tD \Omega^{10(32)} = \partial_\tD^2 \Omega^{10(32)} = 0$ at $\tD = 0$. At these configurations, low dephasing due to fluctuations of $\tD$ can be expected \cite{chirolli_decoherence_2008, benito_electric-field_2019}. For the system parameters given in Table~\ref{tab:system paramets}, the second-order sweet spot of the lower doublet is found at $\theta = 0.35\pi$ [Fig.~\ref{fig:Heff_spectrum}(c)] and for the upper doublet at $\theta = 0.4\pi$.

Because of the spin-charge hybridization being present for all $|B_x| >0$, the intra-doublet transitions ($\ket{0} \leftrightarrow \ket{1}$ and $\ket{2} \leftrightarrow \ket{3}$) couple to photons for all $\theta > 0$. In the case of small $\theta$ and $\delta \ll 2\alpha l$, the coupling is between states with (almost) opposite spin $\braket{\sigma_z} \approx \pm 1$ and angular momentum $\braket{L} \approx \pm \hbar l$, whereas for $\theta = \pi/2$, the SOC hybridizes bonding and antibonding orbitals with the spin in the $x$-direction which is analogous to the flopping-mode mechanism known from conventional DQDs \cite{benito_input-output_2017, benito_electric-field_2019, croot_flopping-mode_2020}.

The highly anisotropic magnetic field dependence of the system leads to multiple interesting operational configurations. The spin-photon coupling is maximized for $\theta = \pi/2$ and $\tD = 0$. On the other hand, operating at a second-order sweet spot somewhat reduces the spin-photon coupling strength, but is expected to dramatically reduce dephasing. 
Finally, the DQD can be electrically detuned from the even-odd crossing using $\tD \gg 0$, which results in spin doublets of single QD states with disabled spin-photon coupling, and therefore also reduced decoherence due to charge noise.
This enables a high degree of control of the system properties using the magnetic field and electric detuning.

\subsection{Transitions in the Kramers doublet regime} \label{sec:2lvl}

In the following, we focus on the small magnetic field regime with $|\mu_B g \bm{B}|, \mu^*_B l B_z \ll \epsilon$. In this regime, although the Kramers degeneracy of the doublets is lifted, the time-reversal symmetry of the doublet states is approximately conserved. In other words, the small magnetic field mainly causes a mixing of the intra-doublet states while mixing of the upper and lower doublet is negligible. As described in Appendix~\ref{app:smallField}, this allows us to consider the upper and lower doublet separately which leads to analytical expressions for the doublet splitting and photon coupling strengths of the intra-doublet transitions. In general, any two states can be coupled by microwave photons. A short discussion is given in Appendix~\ref{app:transitions}.

In the small field regime, the Kramers doublets are described by the effective two-level spin Hamiltonian
\begin{align}
    &H_\mathrm{small}^\pm
    = \frac{\mu_B}{2}\left(\begin{matrix}
    g^{\pm}_z B_z & A^\mp g B_x  \\
    A^\pm g B_x &  - g^{\pm}_z B_z  
\end{matrix}\right), \label{eq:Hsmall}
\end{align}
where the $\pm$ superscript denotes the lower (-) and upper (+) doublet and the basis states are defined in Eqs.~\eqref{eq:zeroB_basis_0}-\eqref{eq:zeroB_basis_3}. These feature separable orbital and spin degrees of freedom with spin quantization in $z$. For $\delta = \tD = 0$, the doublets consist of eigenstates of $\Sigma_y$, i.e. ring states, with opposite angular momentum, whereas for finite $\delta$ and $\tD$, $|\braket{\Sigma_y}|$ is reduced and $|\braket{\Sigma_x}|, |\braket{\Sigma_z}| > 0$.
The $g$-factor in the $z$-direction is given by
\begin{align}
    g^{\pm}_z = g \mp \frac{4 \alpha l^2}{\epsilon}\frac{m}{m^*}, \label{eq:gz}
\end{align}
and the $g$-factor in the $x$-direction is modulated by the complex factor
\begin{align}
    A^\pm = \frac{\pm 2i\alpha l \tD/\epsilon -\delta}{\sqrt{4\alpha^2 l^2 + \delta^2}}\,.
\end{align}
For $\delta = 0$ and $\tD=0$, $A^\pm = 0$ which means that the $g$-factor in the $x$-direction is fully suppressed and the $\Sigma_y, \sigma_z$ quantization is approximately conserved for small $\bm{B}$-fields.

Figures~\ref{fig:small_field}(a),(b) display the splittings $\Omega^\pm = \mu_B \sqrt{|A^\pm|^2g^2 B_x^2 + (g_z^\pm)^2 B_z^2}$ as a function of $\tD$ and the angle of the magnetic field $\theta$ for the upper and lower doublets. Due to the opposite spin and orbital contributions to $g_z^-$, for $\theta = 0$, the lower doublet states cross at $g_z^- = 0$, i.e.,
\begin{align}
    \tD =& \pm \sqrt{4 \alpha^{2} l^{2}\left(4\frac{l^{2} m^2}{{g}^{2} {m^*}^2} - 1\right) - \delta^{2}}. \label{eq:Bz_cross}
\end{align}
 Contrarily, for the upper doublet, the orbital and spin components of $g_z^+$ are reinforcing each other, which, for $\theta < \pi/2$, results in an overall higher level splitting than for the lower doublet. 

The $\tD$ dependency of $A^\pm$ and $g_z^\pm$ results in a $\tD$-dependent rotation of the eigenvectors of $H_\mathrm{small}^\pm$ and thereby enables the driving of transitions.
The resulting transversal coupling strengths $\gamma_\perp^\pm$ are displayed in Figs.~\ref{fig:small_field}(c), (d) as functions of $\theta$ and $\tD$. 
At $\theta = 0$ the coupling strength $\gamma_\perp^\pm = 0$ because the doublet states possess opposite spin directions without spin-charge hybridization. 
For the lower doublet [Fig.~\ref{fig:small_field}(c)], two branches of maximum coupling strength emerge for small $\theta>0$. These originate from the degenerate level crossings for $\theta = 0$ [see Eq.~\eqref{eq:Bz_cross}] which become avoided level crossings in the presence of an in-plane magnetic field component $B_x$.
For a sufficiently large $\theta$, both branches merge and at $\theta = \pi/2, \tD=0$ the upper and lower doublets reach  their maximum coupling strength of
\begin{align} \label{eq:gamma_perp_x}
    \gamma_\perp^x = \epsilon_\gamma\zeta \frac{\alpha l}{4\alpha^2 l^2 + \delta^2}  \mu_B g B_x,
\end{align}
with a doublet splitting of
\begin{align} \label{eq:Omega_x}
    \Omega^x = \frac{\delta}{\sqrt{4\alpha^2 l^2 + \delta^2}} \mu_B g B_x.
\end{align}
Here one has to be cautious: These equations are only valid if $\mu_B g B_x \ll \sqrt{4\alpha^2 l^2 + \delta^2}$. For high magnetic fields, $\gamma_\perp^\pm$ is limited by the charge qubit coupling strength $\gamma_\perp^\mathrm{cq} = \epsilon_\gamma \zeta/2$ and $\Omega^\mathrm{cq} = \delta$ \cite{childress_mesoscopic_2004}.
While for $\theta = \pi/2$, the splitting and transition rates are the same for the upper and lower doublets,
for $\theta < \pi/2$ higher ratios between $\gamma_\perp^\pm$ and $\Omega_\pm$ can be reached for the lower doublet due to its lower $g$-factor in the $z$-direction. This might makes the lower doublet advantageous for the use as a spin-photon interface. 

\subsection{Charge Noise Sweet Spots} \label{sec:sweet_spot}

In this section, we investigate the aforementioned charge noise sweet spots (see Sec.~\ref{sec:spectrum}) in more detail using the spin Hamiltonian derived in Sec.~\ref{sec:2lvl}.
Arbitrary fluctuation of the electrostatic potential $V(\varphi)$ results in fluctuations of the system parameters $\tD$ and $\delta$, depending on if the fluctuation conserves ($\tD$) or breaks ($\delta$) the parity symmetry $\Pi$, see Appendix~\ref{app:dephasing}. 
Therefore, the system can have separate first and second order charge noise sweet spots of $\tD$ and $\delta$.
A general first-order sweet spot is reached for a configuration of simultaneous sweet spots of $\tD$ and $\delta$, i.e., 
$\partial_\tD \Omega^\pm = \partial_\delta \Omega^\pm  = 0$.
A general second-order sweet spot of both $\tD$ and $\delta$ requires $\partial_\delta \partial_\tD \Omega^\pm = 0$ additionally to vanishing first and second $\tD$-and $\delta$-derivatives.

Because the cavity couples to the DQD detuning $\tD$ in the same way as the charge noise, the $\tD$-dephasing susceptibility is closely connected to the longitudinal coupling rate, i.e. $\gamma_z^\pm/(\zeta \epsilon_\gamma) = \partial_\tD \Omega^\pm$. Therefore, a first-order sweet spot of $\tD$ is found for $\gamma_z^\pm = 0$.

As derived in Appendix~\ref{app:smallField} and \ref{app:dephasing}, the first-order derivatives of the doublet splitting $\Omega^\pm$ are given by
\begin{align}
    \frac{1}{\tD}\partial_\tD \Omega^\pm &= \frac{1}{\delta}\partial_\delta \Omega^\pm = \frac{2 \alpha l^2 \mu_B}{\Omega^\pm \epsilon^4} (\alpha \mu_B g^2 B_x^2 -  g_z^\pm \mu_B^* \epsilon B_z^2). \label{eq:fst_ord_der}
\end{align}
Figures~\ref{fig:small_field}(e-h) display $\partial_\tD \Omega^\pm$ and $\partial_\delta \Omega^\pm$ as a function of $\tD$ and $\theta$.
Additionally to the trivial sweet spots at $\tD = 0$ and $\delta = 0$, the right-hand side of Eq.~\eqref{eq:fst_ord_der} is set to zero for the magnetic field angle $\theta_{ss}^\pm$ given by
\begin{align}
    \tan(\theta_\mathrm{ss}^\pm)^2 = \bigg|\frac{m}{m^*}\frac{g_z^\pm\epsilon}{g^2\alpha}\bigg|.
\end{align}
This means that for $\theta = \theta_{ss}$ the doublet splitting $\Omega^\pm$ has a general first-order sweet spot for fluctuations in both $\tD$ and $\delta$ [green dashed line in Figs.~\ref{fig:small_field}(e)-(h)].
For the lower doublet [Figs.~\ref{fig:small_field}(e),(g)], $\theta_{ss}^-$ looks similar to a half-circle which emerges out of the crossing points at $\theta = 0$ [Eq.~\eqref{eq:Bz_cross}], while for the upper doublet [Figs.~\ref{fig:small_field}(f),(h)], $\theta_{ss}^+$ appears as two almost vertical lines.

For configurations with $\theta = \theta_{ss}^\pm$ and also $\tD = 0$ ($\delta = 0$), the second derivative $\partial^2_{\tD}\Omega^\pm = 0$ ($\partial^2_{\delta}\Omega^\pm = 0$) and a second-order sweet spot of $\tD$ ($\delta$) is found. 
A general second-order sweet spot of both $\tD$ and $\delta$ is found if, in addition to $\theta = \theta_\mathrm{ss}$, also $\tD = 0$ and $\delta = 0$. For this configuration, the effects of an arbitrary potential fluctuation vanish up to second order. In contrast to the controllable inter-dot detuning $\tD$, it might be harder to achieve $\delta = 0$ because, as shown in Sec.~\ref{sec:eff_cross_H}, $\delta \ne 0$ is caused by symmetry-breaking disorder $V_\delta^o$ in the electrostatic potential.
Nevertheless, by introducing an electric field in the $y$-direction or modifying the barrier strengths of the two barriers independently, the disorder coupling $\delta$ could be suppressed by canceling out the matrix elements $V_\delta^o$ in first-order perturbation theory.

But even for finite $\delta$, the configuration $\tD=0$ and $\theta = \theta_\mathrm{ss}$ leads to a significant reduction of the charge noise susceptibility by being second-order protected against fluctuations of $\tD$ and to first order in fluctuations of $\delta$.
For this configuration, the transition coupling strength is given by
\begin{align} \label{eq:gamma_perp_ss}
    \gamma_\perp^\pm = \epsilon_\gamma\zeta\frac{\alpha l}{4\alpha^2 l^2 + \delta^2}  \mu_B g |\bm{B}|\sin(\theta_{ss}^\pm),
\end{align}
with a doublet splitting of
\begin{align} \label{eq:Omega_ss}
    \Omega^\pm &= \mu_B \sqrt{|A^\pm|^2 g^2  \sin^2(\theta_{ss}^\pm) + (g_z^\pm)^2 \cos^2(\theta_{ss}^\pm)} |\bm{B}|.
\end{align}
Although the ratio between transition rate and splitting $\gamma_\perp^\pm/\Omega^\pm$ is lower than at the maximum driving configuration at $\theta = 0$, [see Eqs.~\eqref{eq:gamma_perp_x}, \eqref{eq:Omega_x}] the reduced charge noise susceptibility, as well as the lower required $B$-field due to $|g_z^\pm| > |A^\pm g|$ might make this configuration the preferable operating point.

\begin{figure}
    \centering
    \includegraphics[width=8.4cm]{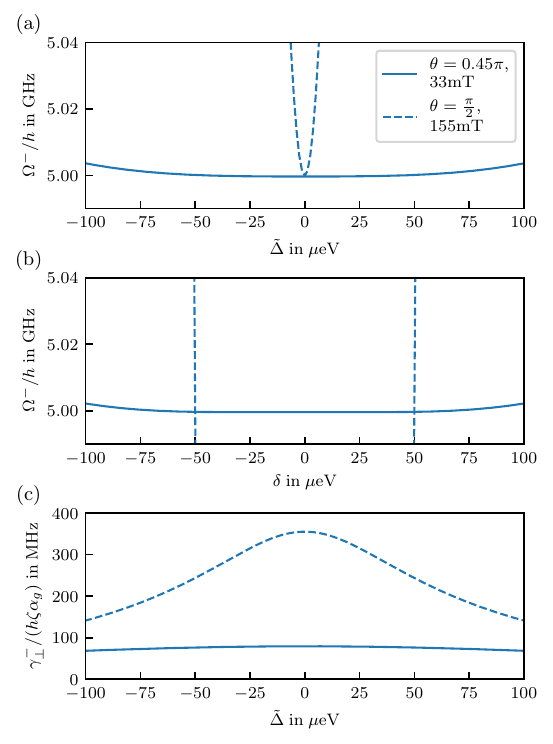}
    \caption{(a) Level splitting $\Omega^-$ of the lower doublet as a function of $\tD$ for parameters of a circular DQD in an InAs nanowire taken from Ref.~\cite{potts_electrical_2019}. The solid line is the second-order sweet spot configuration with $\theta =\theta_{ss}^- = 0.45\pi$ and $|\bm{B}| = 33\,\mathrm{mT}$ and the dashed line shows the maximum coupling configuration with $\theta = \pi/2$ and $|\bm{B}| = 155\,\mathrm{mT}$. (b) $\Omega^-$ as a function of $\delta$ for the same configurations. Here, we assume fixed $\tD = 0$ while $\delta$ is fixed at $50\,\mathrm{\mu eV}$ for the other plots.
    (c) Coupling strength $\gamma_\perp^-$ as a function of $\tD$ divided by the cavity-DQD lever arm $\zeta \alpha_g$. We assume a resonator impedance $Z = 1\,\mathrm{k\Omega}$ and cavity resonance frequency of $\omega=5\,\mathrm{GHz}$.}
    \label{fig:InAs_example}
\end{figure}

\subsection{Experimental Consideration} \label{sec:InAs}

We use the measurement results from Potts et al. \cite{potts_electrical_2019} on a circular DQD system in an InAs nanowire to extract example parameters for our model. The parameters $\delta = 50\,\mathrm{\mu eV}$ and $\alpha l = 117\,\mathrm{\mu eV}$ were derived from electronic transport measurements and for the electron $g$-factor, the bulk InAs value of $g=-12$ is used.
Using the effective four-level Hamiltonian $H_\mathrm{eff}$, Eq.~\eqref{eq:H_4lvl}, we can calculate the required magnetic field configuration for a fixed level splitting of $\Omega^-/h = 5\,\mathrm{GHz}$ between the lower doublet states. 
Figures~\ref{fig:InAs_example}(a)-(c) compare the obtained $\Omega^-$ and coupling rates $\gamma_\perp^-$ of the maximum coupling configuration $\theta = \pi/2$ with $|\bm{B}| = 155\,\mathrm{mT}$ with the second order sweet spot configuration which is found to be at $\theta_{ss}^- \approx 0.45\pi$ with $|\bm{B}| = 33\,\mathrm{mT}$. 
In addition to a lower required magnetic field strength, it is clearly visible that $\Omega^-$ in the second-order sweet spot configuration is much less sensitive to fluctuations in $\tD$ [Fig.~\ref{fig:InAs_example}(a)] and $\delta$ [Fig.~\ref{fig:InAs_example}(b)]. 
To calculate the transition rate $\gamma_\perp^-$ in Fig.~\ref{fig:InAs_example}(c), we assumed a high-impedance cavity with $Z=1\,\mathrm{k\Omega}$ and resonance frequency $\omega = 5\,\mathrm{GHz}$. The maximum coupling strength for $\theta = \pi/2$ is given by $\gamma_\perp^-/(h \zeta \alpha_g) = 355\,\mathrm{MHz}$ and for $\theta = 0.45\pi$ by $\gamma_\perp^-/(h \zeta \alpha_g) = 79\,\mathrm{MHz}$. For realistic devices, the coupling strengths are reduced by lever arms $\zeta \alpha_g < 1$.

A higher coupling strength $\gamma_\perp^-$ at the second order $\tD$ sweet spot for a fixed splitting $\Omega^-/h = 5\,\mathrm{GHz}$ is possible for smaller $g_z^-$, i.e., smaller angular momentum $l$: in this case, the required larger $B_x$ field and lower SOC lead to more charge-like doublets with higher coupling strength. 
On the other hand, reducing $l$, also makes the eigenstates less ring-like. To ensure the ring nature of the orbitals, $\delta \ll \alpha l$ should be satisfied.

\section{Conclusion} \label{sec:conclusion}

We have proposed and analytically investigated a novel platform for spin-photon interfaces based on a circular DQD geometry. Our effective ring model captures the interplay between spin-orbit coupling, magnetic flux and disorder, and shows the emergence of ring states. We have shown how a tilted magnetic field introduces spin-charge hybridization and therefore enables spin-photon coupling.
A key result is the identification of a second-order charge noise sweet spot at a specific magnetic field angle, where the spin-photon coupling remains strong and the system is robust to low-frequency electrical noise. This sweet spot enables operation with reduced decoherence while preserving controllability, and the coupling can be switched off dynamically by detuning the QD levels or rotating the magnetic field. Using realistic material parameters, extracted from InAs nanowire experiments, we predict that coupling rates of tens to hundreds of $\mathrm{MHz}$ can be achieved at relatively small magnetic fields.

Our findings suggest that quantum ring-based DQD systems offer a versatile and noise-resilient platform for coherent spin-photon interfaces, with promising applications in quantum information processing and hybrid quantum systems.

\section{Acknowledgments}

We thank Peter Samuelsson for his ideas and insightful discussions, and the Knut and
Alice Wallenberg Foundation through the Wallenberg Center for Quantum Technology (WACQT) and NanoLund for funding. FVB acknowledges funding from the Swedish Research Council (VR) and from the European Union’s Horizon Europe research and innovation programme under the Marie Sk{\l}odowska-Curie grant agreement No~101204715.
\\

\appendix

\section{Continuity Condition} \label{sec:cont_cond}

The ansatz in Eqs.~\eqref{eq:wavefunc_even} and \eqref{eq:wavefunc_odd} diagonalizes the zero-field Hamiltonian, Eq.~\eqref{eq:H0}, if the continuity conditions at the Dirac barriers
\begin{align}
    n \mathrm{~even}:~& A_j \cos \left(k^+_j\frac{\pi}{2} \right) = B_j \cos \left(k^-_j \frac{\pi}{2} \right),\\
    n \mathrm{~odd}:~& C_j \sin \left(k^+_j\frac{\pi}{2}\right) = -D_j \sin \left(k^-_j \frac{\pi}{2} \right),
\end{align}
are fulfilled. Additionally, the derivatives must also be continuous
\begin{align}
    n \mathrm{~even}:~&- \frac{\beta}{\epsilon_R} +  k^{-}_n \tan{\left(\frac{\pi k^{-}_n}{2} \right)} +  k^{+}_n \tan{\left(\frac{\pi k^{+}_n}{2} \right)} = 0\,, \label{eq:Ce}\\
    n \mathrm{~odd}:~&\frac{\beta}{\epsilon_R} + k^{-}_n \cot{\left(\frac{\pi k^{-}_n}{2} \right)} + k^+_n \cot{\left(\frac{\pi k^{+}_n}{2} \right)} = 0\,. \label{eq:Co}
\end{align}

\section{Small Field Spectrum} \label{app:smallField}

Without a magnetic field, the energy spectrum  of $H_\mathrm{eff}$, Eq.~\eqref{eq:H_4lvl}, consists of two degenerate Kramers doublets which we denote with $+$ for the upper and $-$ for the lower. We investigate the following Kramers basis

\begin{align} 
&\ket{-, \uparrow} = \frac{1}{\sqrt{2}}\left(-e^{i\phi/2}c_- \ket{e} + e^{-i\phi/2}c_+ \ket{o} \right)\ket{\uparrow}, \label{eq:zeroB_basis_0}\\
&\ket{-, \downarrow} = \frac{1}{\sqrt{2}}\left(e^{-i\phi/2}c_- \ket{e} - e^{i\phi/2}c_+ \ket{o} \right)\ket{\downarrow},\label{eq:zeroB_basis_1}\\
&\ket{+, \uparrow} = \frac{1}{\sqrt{2}}\left(e^{i\phi/2}c_+ \ket{e} + e^{-i\phi/2}c_- \ket{o} \right)\ket{\uparrow}, \label{eq:zeroB_basis_2}\\
&\ket{+, \downarrow} = \frac{1}{\sqrt{2}}\left(-e^{-i\phi/2}c_+ \ket{e}- e^{i\phi/2}c_-\ket{o}\right)\ket{\downarrow}, \label{eq:zeroB_basis_3}
\end{align}
where
\begin{align}
    c_\pm = \sqrt{1 \pm \frac{\tD}{\epsilon}}, \quad \phi = \arctan\left(\frac{2\alpha l}{\delta}\right).
\end{align}
For $\delta \ll \alpha l$, the upper and lower doublet states are eigenstates of $\Sigma_y$, i.e., ring-states with opposite angular momentum, whereas for increasing $\delta$, the states become less ring-like, i.e., $\braket{\Sigma_y}$ decreases and $\braket{\Sigma_x}$ increases.
In this basis, $H_\mathrm{eff}$ is given by
\begin{widetext}
    \begin{align}
&H_\mathrm{eff}
= \left(\begin{matrix}
-\frac{\epsilon}{2} + \frac{\mu_B g^{-}_z}{2}B_z & A^+\frac{\mu_B g}{2}B_x & iA^+\mu_B^* lB_z & i\frac{\mu_B g \alpha l}{\epsilon} B_x \\
A^- \frac{\mu_B g}{2}B_x & -\frac{\epsilon}{2} - \frac{\mu_B g^{-}_z}{2}B_z & -i\frac{\mu_B g \alpha l}{\epsilon} B_x & iA^-\mu_B^* lB_z \\
-iA^-\mu_B^* lB_z & i\frac{\mu_B g \alpha l}{\epsilon} B_x & \frac{\epsilon}{2} + \frac{\mu_B g^{+}_z}{2}B_z & A^- \frac{\mu_B g}{2}B_x \\
-i\frac{\mu_B g \alpha l}{\epsilon} B_x & iA^+\mu_B^* lB_z & A^+ \frac{\mu_B g}{2}B_x & \frac{\epsilon}{2} - \frac{\mu_B g^{+}_z}{2}B_z 
\end{matrix}\right), \label{eq:Heff_matrix}
\end{align}
\end{widetext}
where the $g$-factor in the $z$-direction is given by
\begin{align}
    g^{\pm}_z = g \mp \frac{4 \alpha l^2}{\epsilon}\frac{m}{m^*} 
\end{align}
and
\begin{align}
    A^\pm = \frac{\pm 2i\alpha l\frac{\tD}{\epsilon} -\delta}{\sqrt{4\alpha^2 l^2 + \delta^2}}\,.
\end{align}
For $\bm{B} = 0$, Eq.~\eqref{eq:Heff_matrix} is diagonal and the degenerate upper and lower Kramers doublets are split by $\epsilon$ [Eq.~\eqref{eq:epsilon}]. 
In the small field limit $\mu_B g |\bm{B}|, \mu^*_B l B_z \ll \epsilon$, we can neglect the off-diagonal terms between the lower and upper Kramers doublet and arrive at the small field Hamiltonian
\begin{align}
    &H_\mathrm{small}^\pm
    = \frac{\mu_B}{2}\left(\begin{matrix}
    g^{\pm}_z B_z & A^\mp g B_x  \\
    A^\pm g B_x &  - g^{\pm}_z B_z  
\end{matrix}\right), \label{eq:Hsmallapp}
\end{align}
where the $\pm$ superscript of $H_\mathrm{small}^\pm$ also stands for the upper ($+$) and lower ($-$) doublet. 
Using this, the doublet splitting is given by
\begin{align}
    \Omega^\pm = \mu_B \sqrt{|A^\pm|^2g^2 B_x^2 + (g_z^\pm)^2 B_z^2}.
\end{align}
To calculate the Rabi coupling strengths, we write $H_\mathrm{small}^\pm$ in terms of Pauli operators $\tilde{\mathcal{Z}}_\pm = \ket{\pm,\uparrow} \bra{\pm, \downarrow} - \ket{\pm,\downarrow} \bra{\pm, \uparrow}, \tilde{\mathcal{X}}_\pm = \dots$~ using the Kramers basis states and obtain
\begin{align}
    &H_\mathrm{small}^\pm = h_x^\pm \tilde{\mathcal{X}}_\pm + h_y^\pm \tilde{\mathcal{Y}}_\pm + h_z^\pm \tilde{\mathcal{Z}}_\pm
    \\&= - \frac{\mu_B g B_x}{2\sqrt{4\alpha^2 l^2 + \delta^2
    }}( \delta\tilde{\mathcal{X}}_\pm \mp \frac{2\alpha l\tD}{\epsilon}\tilde{\mathcal{Y}}_\pm) + \frac{\mu_B g_z^\pm B_z}{2}\tilde{\mathcal{Z}}_\pm.
\end{align}
As described in Sec.~\ref{sec:cavity_coupl}, by calculating the $\Delta$-derivative of the Hamiltonian
\begin{align}
    &(H_{2x2}^{\pm})' = (h_x^\pm)' \tilde{\mathcal{X}}_\pm + (h_y^\pm)' \tilde{\mathcal{Y}}_\pm + (h_z^\pm)' \tilde{\mathcal{Z}}_\pm \\
    &=\pm\frac{\zeta\alpha l}{\epsilon^3}\left(\sqrt{4\alpha^2 l^2 + \delta^2}  \mu_B g B_x \tilde{\mathcal{Y}}_\pm + 2\tD \mu_B^* l B_z\tilde{\mathcal{Z}}_\pm\right),
\end{align}
we can derive the coupling strengths of the Rabi Hamiltonian, Eq.~\eqref{eq:H_omega}. Here, $\zeta = \partial_\Delta \tD$. 
The longitudinal Rabi coupling is given by the $\Delta$-derivative of $\Omega^\pm$ which is the same as the parallel component of $(\bm{h}^\pm)'$ to $\bm{h}^\pm$
\begin{align}
    \gamma_z^\pm &= \frac{1}{2}\epsilon_\gamma \zeta \partial_\tD \Omega^\pm\\
    &= 2 \epsilon_\gamma\frac{h_y^\pm (h^\pm_y)' + h_z^\pm (h^\pm_z)'}{\Omega^\pm}\\ 
    &= \zeta\epsilon_\gamma \frac{2\tD \alpha l^2 \mu_B}{\Omega^\pm \epsilon^4} (\alpha \mu_B g^2 B_x^2 -  g_z^\pm \mu_B^* \epsilon B_z^2). \label{eq:gamma_z_2lvl}
\end{align}
By calculating the orthogonal component of $(\bm{h}^\pm)'$ to $\bm{h}^\pm$, the transversal coupling strength is found,
\begin{align}
    \gamma_\perp^\pm = \sqrt{\epsilon_\gamma^2(h_y')^2 + \epsilon_\gamma^2(h_z')^2 - \gamma_z^2}.
\end{align}
For $\tD = 0$, it is given by
\begin{align}
    \gamma_\perp^\pm = \epsilon_\gamma\zeta \frac{\alpha l}{4\alpha^2 l^2 + \delta^2}  \mu_B g B_x.
\end{align}

\begin{figure*}
    \centering
    \includegraphics[width=\textwidth]{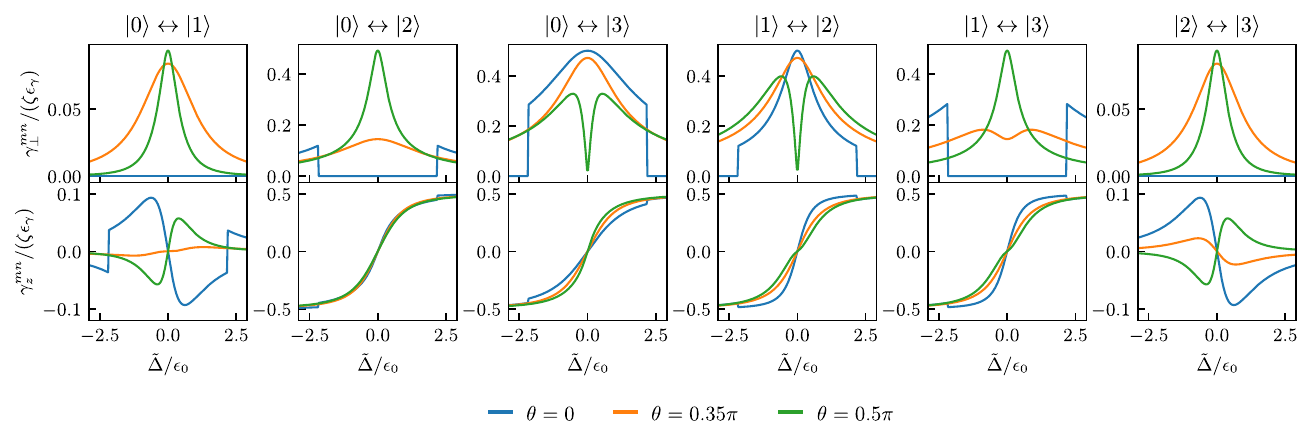}
    \caption{Coupling rates $\gamma_\perp^{mn}, \gamma_{z}^{mn}$ for the transitions $\ket{m} \leftrightarrow \ket{n}$ as a function of the detuning $\tD$. The respective energy spectra are displayed in Figs.~\ref{fig:Heff_spectrum} (b), (c), (d). The jumps in the $\theta = 0$ spectrum are caused by a crossing of the $\ket{0}$ and $\ket{1}$ states close to $\tD = 0$. For $\theta = 0$, only transitions between same spin states are allowed, therefore $\gamma_\perp^{01} = \gamma_\perp^{02} = \gamma_\perp^{13} = \gamma_\perp^{23} = 0$ close to $\tD = 0$. For $\theta = 0.35\pi$, all transitions are allowed and the $\ket{0} \leftrightarrow \ket{1}$ transition has a second-order sweet spot at $\tD = 0$. The system parameters used for the plots are given in Tab.~\ref{tab:system paramets}.}
    \label{fig:all_couplings}
\end{figure*}

\section{Transitions} \label{app:transitions}

Figure~\ref{fig:all_couplings} displays the Rabi coupling strengths $\gamma_\perp^{mn}$ and $\gamma_z^{mn}$ as a function of $\tD$ for the spectra displayed in Figs.~\ref{fig:Heff_spectrum}(b)-(d). The jumps in the $\theta = 0$ plots are caused by the inversion of the $\ket{0}$ and $\ket{1}$ states. For $\theta = 0$, only transitions between same spin states are allowed; therefore $\gamma_\perp^{01} = \gamma_\perp^{02} = \gamma_\perp^{13} = \gamma_\perp^{23} = 0$ close to $\tD = 0$.
For $\theta = 0.35\pi$, all transitions are allowed, but for $\theta =0.5\pi$ the transition rates $\gamma_\perp^{03} = \gamma_\perp^{12} = 0$ at $\tD = 0$.

\section{Relaxation and Dephasing} \label{app:dephasing}

We model charge noise with an arbitrary noise potential $\delta V(\varphi)$ which can be divided into an even and odd part with respect to the geometrical parity symmetry $\Pi$, i.e., $\delta V = \delta V_e + \delta V_o$. Using Eqs.~\eqref{eq:wavefunc_even}, \eqref{eq:wavefunc_odd} and \eqref{eq:H_4lvl}, we obtain on the subspace of the crossing ($\ket{e}, \ket{o}$)
\begin{align}
    &\delta V_e = \frac{\delta v_e}{2}\Sigma_z  = \delta v_e \partial_\tD H_\mathrm{eff} \label{eq:even_fluc}\\
    &\delta V_o = \frac{\delta v_o}{2}\Sigma_x  = \delta v_o \partial_\delta H_\mathrm{eff}
\end{align}
where $\braket{e|\delta V_e|e} - \braket{o|\delta V_e|o} = \delta v_e$ and $\braket{e|\delta V_o|o} = \delta v_o/2$.
By comparing with $H_\mathrm{eff}$, Eq.~\eqref{eq:H_4lvl}, one sees that considering the potential fluctuation $\delta V$ is equivalent with substituting $\tD \rightarrow \tD + \delta v_e$ and $\delta \rightarrow \delta + \delta v_o$.
Therefore, even fluctuations $\delta V_e$ cause fluctuations of the inter-dot detuning $\tD$ and odd fluctuations $\delta V_o$ cause fluctuations of the disorder coupling $\delta$. 

In the following, we consider the effect of these fluctuations on a qubit consisting of the states $\ket{0}$ and $\ket{1}$ for which $\mathcal{X}, \mathcal{Y}$ and $\mathcal{Z}$ are the respective Pauli matrices.
We define the decoherence Hamiltonian $H_d$ as \cite{chirolli_decoherence_2008, benito_electric-field_2019}
\begin{align}
    H_d = \frac{1}{2}\delta \Omega \mathcal{Z} + \delta \Omega_\perp \mathcal{X},
\end{align}
where the longitudinal fluctuation of the qubit splitting $\delta \Omega$ causes dephasing and the transversal fluctuation $\delta \Omega_\perp$ results in relaxation. Up to first order in $\delta V$, the transversal fluctuation is given by
\begin{align}
    \delta \Omega_\perp &= |\braket{1|\delta V|0}| \\
    &= \frac{1}{2}|\delta v_e \braket{1|\Sigma_z|0} + \delta v_o \braket{1|\Sigma_x|0}|,
\end{align}
and to second order, the longitudinal fluctuation is given by
\begin{align}
    \delta \Omega &= \delta v_e \partial_\tD \Omega + \delta v_o \partial_\delta\Omega \nonumber\\
    &+ \frac{1}{2}\delta v_e^2 \partial_\tD^2 \Omega + \frac{1}{2}\delta v_o^2 \partial_\delta^2 \Omega + \delta v_e\delta v_o\partial_\tD \partial_\delta \Omega \\
    &= \braket{1|\delta V|1} - \braket{0|\delta V|0} + \frac{|\braket{0|\delta V|1}|^2}{\Omega}\nonumber \\
   &+ \frac{1}{2}\sum_{k = 2,3} \left(\frac{|\braket{k|\delta V|1}|^2}{E_1 - E_k} - \frac{|\braket{k|\delta V|0}|^2}{E_0 - E_k}\right),
\end{align}
with the qubit splitting $\Omega = E_1 - E_0$. 

By comparing Eq.~\eqref{eq:even_fluc} with Eq.~\eqref{eq:long_coupl} and Eq.~\eqref{eq:trans_coupl}, we see that the first-order relaxation and dephasing terms caused by the even potential fluctuation $\delta V_e$ are proportional to the Rabi coupling rates $\gamma_\perp$ and $\gamma_z$, i.e., $|\braket{1|\delta V_e|0}| \propto \gamma_\perp^{10}$ and $\braket{1|\delta V_e|1} - \braket{0|\delta V_e|0} \propto \gamma_z^{10}$.  

While dephasing is mainly caused by low frequency noise, relaxation is mainly caused by noise with a frequency close to $\Omega$ \cite{chirolli_decoherence_2008}. 
Independent of the noise spectrum, relaxation and dephasing can be reduced by decreasing the system's susceptibility to the fluctuations, i.e. by reducing the matrix elements $\braket{1|\Sigma_{x,z}|0}$ and the derivatives of the splitting $\partial_\tD \Omega, \partial_\delta\Omega, \partial_\tD^2 \Omega, \partial_\delta^2 \Omega$ and $\partial_\tD \partial_\delta \Omega$. 
In the small field regime, the $\tD$-derivatives are given by Eq.~\eqref{eq:gamma_z_2lvl} with $2\gamma_z/(\zeta \epsilon_\gamma) = \partial_\tD \Omega$. Using the small field approximation from Appendix~\ref{app:smallField}, the $\delta$-derivative can be calculated analogously to $\partial_\tD \Omega$ and is given by
\begin{align}
    \partial_\delta \Omega^\pm = \frac{2\delta \alpha l^2 \mu_B}{\Omega^\pm \epsilon^4} (\alpha \mu_B g^2 B_x^2 -  g_z^\pm \mu_B^* \epsilon B_z^2).
\end{align}

\bibliography{literatur}

\end{document}